\documentstyle[amssymb,prl,aps,twocolumn,graphics]{revtex}

\begin{document}
\draft
\title{Nonperturbative Coherent Population Trapping: An Analytic Model}
\author{V. Delgado and J. M. Gomez Llorente}
\address{Departamento de F\'\i sica Fundamental II,\\
Universidad de La Laguna, 38205-La Laguna, Tenerife, Spain}
\maketitle

\begin{abstract}
Coherent population trapping is shown to occur in a driven symmetric
double-well potential in the strong-field regime. The system parameters have
been chosen to reproduce the $0^{-}\leftrightarrow 3^{+}$ transition of the
inversion mode of the ammonia molecule. For a molecule initially prepared in
its lower doublet we find that, under certain circumstances, the $3^{+}$
level remains unpopulated, and this occurs in spite of the fact that the
laser field is resonant with the $0^{-}\leftrightarrow 3^{+}$ transition and
intense enough so as to strongly mix the $0^{+}$ and $0^{-}$ ground states.
This counterintuitive result constitutes a coherent population trapping
phenomenon of nonperturbative origin which cannot be accounted for with the
usual models. We propose an analytic nonperturbative model which accounts
correctly for the observed phenomenon.
\end{abstract}

\pacs{42.50.Hz, 42.50.Gy, 03.65.-w}

Quantum dynamics in symmetric double-well potentials is important to
understand numerous physical and chemical processes. A typical example is
the tunneling dynamics of the hydrogen atoms in the inversion mode of the
ammonia molecule, which is responsible for the splitting of the vibrational
levels \cite{Hund}. Other examples include electron tunneling in quantum
semiconductor structures \cite{Holt1} or intermolecular proton transfer
processes \cite{Oppen}.

In recent years there has been increasing interest in quantum coherence
phenomena displayed by atomic and molecular systems irradiated with strong
laser fields \cite{Fic1}. Coherent external fields induce quantum
interference effects such as coherent population trapping \cite{Ari1},
electromagnetically induced transparency \cite{Harr1} or lasing without
inversion \cite{Koch1}. In particular, in connection with the time evolution
of a quantum system in a symmetric double-well potential it has been shown
that under certain circumstances an intense laser field can induce coherent
tunneling suppression \cite{Gros1}. In this Letter we show that such a
system can also exhibit coherent population trapping. This population
trapping phenomenon is nonperturbative in nature and cannot be accounted for
with the usual models. We propose an analytically solvable nonperturbative
model which accounts correctly for the essential features of the observed
phenomenon.

Specifically, we consider a symmetric quartic double-well potential strongly
driven by a linearly polarized laser field. After appropriate scaling the
corresponding dimensionless Hamiltonian reads

\begin{equation}
H=\frac{P^{2}}{2}-\frac{X^{2}}{4}+\frac{X^{4}}{64\alpha }-\lambda X\cos
\left( \tau \right) ,  \label{ec1.1}
\end{equation}
where the coupling constant $\lambda $ is proportional to the laser field
amplitude ${\bf E}_{0}$ and $\tau =\omega _{{\rm {L}}}t$ with $\omega _{{\rm 
{L}}}$ being the laser frequency. The dimensionless parameter $\alpha $,
which gives approximately the number of doublets below the barrier top, has
been chosen to be $1.735.$ This value reproduces, to a good approximation,
the effective potential involved in the inversion mode of the ammonia
molecule. The laser frequency has been tuned to the $0^{-}\leftrightarrow
3^{+}$ vibrational transition and its intensity satisfies $\lambda \langle
0^{+}\left| X\right| 0^{-}\rangle ={\bf E}_{0}{\bf \mu }_{12}=0.35\pi \omega
_{{\rm {L}}}$ where ${\bf \mu }_{12}$ is the dipole matrix element between
the ground states $|0^{+}\rangle $ and $|0^{-}\rangle $, and all quantities
are assumed to be dimensionless.

Transitions $0^{+}\leftrightarrow 0^{-}$ and $0^{-}\leftrightarrow 3^{+}$
are dipole allowed whereas the $0^{+}\leftrightarrow 3^{+}$ transition is
forbidden. Therefore, in the weak-field regime (${\bf E}_{0}{\bf \mu }%
_{12}/\omega _{{\rm {L}}},\;\Delta _{0}/\omega _{{\rm {L}}}\ll 1$ with $%
\Delta _{0}$ being the energy splitting of the lower doublet) and for a
laser field tuned to the $0^{-}\leftrightarrow 3^{+}$ transition, one
expects the upper level to be populated or not depending on whether the
molecule is initially prepared in the $|0^{-}\rangle $ or $|0^{+}\rangle $
state, respectively. The laser intensity considered above, however,
corresponds to the strong-field regime (${\bf E}_{0}{\bf \mu }_{12}\approx
\omega _{{\rm {L}}}$). Under these circumstances the two lower levels become
strongly mixed and the $|0^{-}\rangle $ state becomes highly populated. One
then would expect the upper level to be populated irrespective of the fact
that the molecule be initially prepared in the $|0^{-}\rangle $ or $%
|0^{+}\rangle $ state.

Fig. 1a shows the evolution of the populations for an ammonia molecule
initially prepared in its ground state $|0^{+}\rangle $. These results have
been obtained numerically by direct integration of the Schr\"{o}dinger
equation. 
We have included the 20 lowest-lying levels, which guarantees convergence.
As is apparent from the figure, the upper level $|3^{+}\rangle $
remains always unpopulated [curve (3)], and this occurs despite the fact
that the $|0^{-}\rangle $ state becomes highly populated [curve (2)] and the
laser field directly connects this latter state with the upper level. This
figure also shows that the total population in the lower doublet remains
always close to unity [curve (1)]. Thus, under the action of the driving
field the initial population oscillates rapidly between $|0^{+}\rangle $ and 
$|0^{-}\rangle $ while it remains trapped in the lower doublet. This
counterintuitive result represents a coherent population trapping phenomenon
of nonperturbative nature which cannot be accounted for with the usual
models.

The case of an ammonia molecule prepared initially in $|0^{-}\rangle $ is
considered in Fig. 1b. This figure shows the time evolution of both the
population of the upper level $|3^{+}\rangle $ [curve (3)] and the total
population of the lower doublet [curve (1)]. As before the initial
population oscillates very rapidly between $|0^{-}\rangle $ and $%
|0^{+}\rangle $ (not shown for clarity). Now, however, a periodic population
transfer between the lower doublet and the upper level takes place on a
different timescale. In fact, apart from the rapid oscillations of the upper
level population (which originates from population transfers to levels
adjacent to $|3^{+}\rangle $, as a detailed numerical analysis reveals) the
behavior of the system in this nonperturbative regime resembles that of the
corresponding weak-field regime.


\begin{figure}{\par\centering \resizebox{8.2cm}{!}{\rotatebox{0}
{\includegraphics{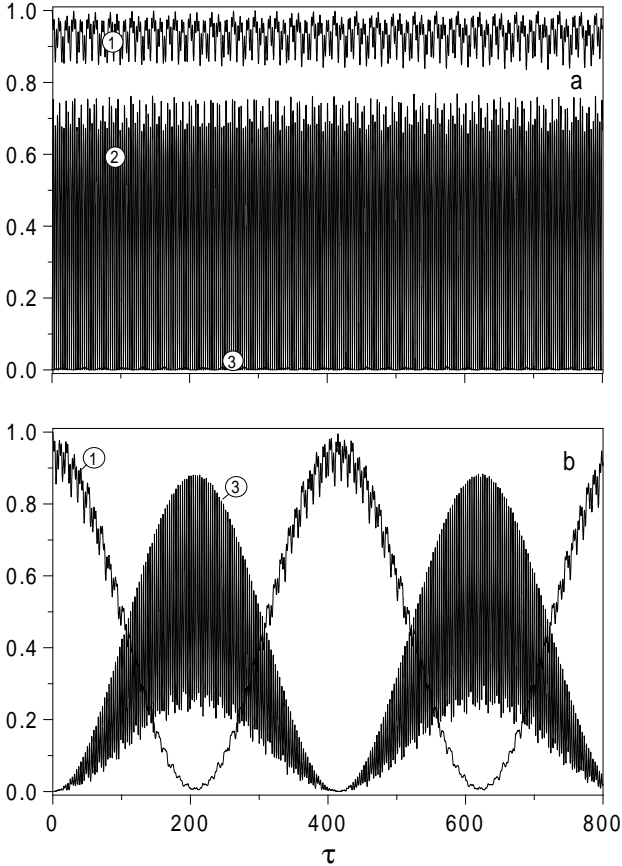}}} \par}
\caption{\small
Populations vs. $\tau =\omega _{{\rm {L}}}t$ for an ammonia molecule
initially prepared in: (a) the $|0^{+}\rangle $ state; and (b) the $%
|0^{-}\rangle $ state. Curves (1) give the total population of the lower
doublet; curve (2) gives the population of the $|0^{-}\rangle $ state; and
curves (3) give the population of the upper level $|3^{+}\rangle $. %
}\end{figure}

The above results pose two intriguing questions: i) Why does the population
become trapped when the molecule is initially prepared in its ground state?
and ii) Why does the system behave essentially in a similar way both in the
nonperturbative strong-field regime and in the weak-field regime? In what
follows, we propose an analytic nonperturbative three-level model which can
give an answer to these questions.

The most directly involved states, $|0^{+}\rangle $, $|0^{-}\rangle $ and $%
|3^{+}\rangle $, will be denoted $|1\rangle $, $|2\rangle $ and $|3\rangle $%
, respectively. The energy splitting of the lower doublet is $\Delta _{0}$,
and $\omega _{3}$ denotes the energy of the upper level. The system
Hamiltonian is

\begin{eqnarray}
H=\frac{\Delta _{0}}{2}\left( \sigma _{22}-\sigma _{11}\right) &+&\omega
_{3}\sigma _{33}-\Omega _{12}\cos \left( \tau \right) \left( \sigma
_{12}+\sigma _{21}\right)  \nonumber \\
&-&\Omega _{23}\cos \left( \tau \right) \left( \sigma _{23}+\sigma
_{32}\right) ,  \label{ec1.2}
\end{eqnarray}
where $\hbar \equiv 1$; $\sigma _{ij}\equiv |i\rangle \langle j|$; and $%
\Omega _{ij}\equiv {\bf E}_{0}{\bf \mu }_{ij}$ with ${\bf \mu }_{ij}$ being
the dipole matrix elements between $|i\rangle $ and $|j\rangle $.

The most rapidly oscillating terms can be absorbed by performing the 
unitary transformation

\begin{equation}
U(\tau )=\exp \left[ -i\frac{\Omega _{12}}{\omega _{{\rm {L}}}}\left( \sigma
_{12}+\sigma _{21}\right) \sin \left( \tau \right) +i\sigma _{33}\tau
\right] ,  \label{ec1.3}
\end{equation}
which leads to the transformed Hamiltonian

\begin{eqnarray}
H^{\prime } &=&\left( \Delta _{0}/2\right) \left\{ \cos \left[ 2\phi (\tau
)\right] \left( \sigma _{22}-\sigma _{11}\right) \right.  \nonumber \\
&+&\left. i\sin \left[ 2\phi (\tau )\right] \left( \sigma _{21}-\sigma
_{12}\right) \right\} +(\omega _{3}-\omega _{{\rm {L}}})\sigma _{33} 
\nonumber \\
&-&\Omega _{23}\cos \left( \tau \right) \left\{ \,e^{-i\tau }\left( \cos
\left[ \phi (\tau )\right] \sigma _{23}\right. \right.  \nonumber \\
&-&\left. \left. i\sin \left[ \phi (\tau )\right] \sigma _{13}\right) +{\rm %
h.c.}\right\} ,  \label{ec1.4}
\end{eqnarray}
with $\phi (\tau )=\left( \Omega _{12}/\omega _{{\rm {L}}}\right) \sin
\left( \tau \right) $. Next, we expand the time dependent coefficients of $%
H^{\prime }$ in Fourier series, which allows us to separate the Hamiltonian
into a dominant constant contribution $H_{0}^{\prime }$ and a time-dependent
part $\Delta H^{\prime }(\tau )$. 
Then, substitution of $H^{\prime }$ into the evolution operator of the 
system shows that when the driving field is quasiresonant
with the $|2\rangle \leftrightarrow |3\rangle $ transition and both the
energy difference $\Delta _{0}$ and the Rabi frequency $\Omega _{23}$ are
small in comparison with the laser frequency, $\Delta H^{\prime }(\tau )$
becomes a small, rapidly oscillating perturbation which can be safely
neglected. More generally, it can be shown that in the strong-field regime ($%
\Omega _{12}/\omega _{{\rm {L}}}\gtrsim 1$) and for a quasiresonant laser
field, $\Delta H^{\prime }(\tau )$ becomes negligible whenever $\Delta
_{0}/\omega _{{\rm {L}}},\;\Omega _{23}/\omega _{{\rm {L}}}\ll \sqrt{\Omega
_{12}/\omega _{{\rm {L}}}}$. 
(In our case, $\Delta_{0}/\omega _{{\rm {L}}} = 3.28 \times 10^{-4},
\Omega _{23}/\omega _{{\rm {L}}} = 0.23$, and 
$\Omega _{12}/\omega _{{\rm {L}}} = 1.10$.)
Under these circumstances, the dynamical evolution of the system is 
governed by the Hamiltonian

\begin{equation}  \label{ec1.5}
H^{\prime }\equiv \frac{\Delta _0^{{\rm R}}}2\left( \sigma _{22}-\sigma
_{11}\right) +(\omega _3-\omega _{{\rm {L}}})\sigma _{33}-\frac{\Omega
_{23}^{{\rm R}}}2\left( \sigma _{23}+\sigma _{32}\right) ,
\end{equation}
where the renormalized energy difference $\Delta _0^{{\rm R}}$ and Rabi
frequency $\Omega _{23}^{{\rm R}}$ are field-dependent quantities defined as 
$\Delta _0^{{\rm R}}=\Delta _0J_0\left( 2\Omega _{12}/\omega _{{\rm {L}}%
}\right) $ and $\Omega _{23}^{{\rm R}}=2\omega _{{\rm {L}}}(\Omega
_{23}/\Omega _{12})J_1\left( \Omega _{12}/\omega _{{\rm {L}}}\right) $, with 
$J_n$ being the n$^{{\rm th}}$-order Bessel function. The Schr\"odinger
equation associated with the above Hamiltonian can be readily solved
analytically, and after transforming back one obtains the following
nonperturbative general solution

\begin{equation}
|\Psi (\tau )\rangle =C_{1}(\tau )|1\rangle +C_{2}(\tau )|2\rangle
+C_{3}(\tau )|3\rangle  \label{ec1.6}
\end{equation}
where

\begin{mathletters}
\begin{equation}  \label{ec1.7}
C_1(\tau )=C_1^{\prime }(\tau )\cos \phi (\tau )+iC_2^{\prime }(\tau )\sin
\phi (\tau )
\end{equation}

\begin{equation}  \label{ec1.8}
C_2(\tau )=C_2^{\prime }(\tau )\cos \phi (\tau )+iC_1^{\prime }(\tau )\sin
\phi (\tau )
\end{equation}

\begin{equation}
C_{3}(\tau )=C_{3}^{\prime }(\tau )e^{-i\tau }  \label{ec1.9}
\end{equation}
and the $C_{i}^{\prime }(\tau )$, which are the probability amplitudes
associated with the Hamiltonian (\ref{ec1.5}), are given by 
\end{mathletters}
\begin{mathletters}
\begin{equation}
C_{1}^{\prime }(\tau )=C_{1}^{\prime }(0)e^{i\frac{\Delta _{0}^{{\rm R}}}{%
2\omega _{{\rm {L}}}}\tau }  \label{ec1.10}
\end{equation}

\begin{eqnarray}
C_{2}^{\prime }(\tau ) &=&\left\{ C_{2}^{\prime }(0)\cos \left( \frac{\Omega
^{{\rm R}}}{2\omega _{{\rm {L}}}}\tau \right) +\frac{i}{\Omega ^{{\rm R}}}%
\left( C_{2}^{\prime }(0)\,\delta ^{{\rm R}}\right. \right.  \nonumber \\
&+& \left. \left. C_{3}^{\prime }(0)\Omega _{23}^{{\rm R}}\right) \sin
\left( \frac{\Omega ^{{\rm R}}}{2\omega _{{\rm {L}}}}\tau \right) \right\}
e^{-\frac{i}{2\omega _{{\rm {L}}}}\left( \delta ^{{\rm R}}+ \Delta _{0}^{%
{\rm R}}\right) \tau }  \label{ec1.11}
\end{eqnarray}

\begin{eqnarray}
C_{3}^{\prime }(\tau ) &=&\left\{ C_{3}^{\prime }(0)\cos \left( \frac{\Omega
^{{\rm R}}}{2\omega _{{\rm {L}}}}\tau \right) -\frac{i}{\Omega ^{{\rm R}}}%
\left( C_{3}^{\prime }(0)\,\delta ^{{\rm R}}\right. \right.  \nonumber \\
&-&\left. \left. C_{2}^{\prime }(0)\Omega _{23}^{{\rm R}}\right) \sin \left( 
\frac{\Omega ^{{\rm R}}}{2\omega _{{\rm {L}}}}\tau \right) \right\} e^{\frac{%
i}{2\omega _{{\rm {L}}}}\left[ \delta ^{{\rm R}}-2\left( \omega _{3}-\omega
_{{\rm {L}}}\right) \right] \tau }  \label{ec1.12}
\end{eqnarray}
where $\delta ^{{\rm R}}=\omega _{3}-\Delta _{0}^{{\rm R}}/2-\omega _{{\rm {L%
}}}$ is the renormalized detuning and $\Omega ^{{\rm R}}=\sqrt{\left( \Omega
_{23}^{{\rm R}}\right) ^{2}+\left( \delta ^{{\rm R}}\right) ^{2}}$ is the
renormalized generalized-Rabi-frequency. The physical content of the above
solution becomes more transparent by considering the extended Hilbert space
of $\tau $-periodic state vectors \cite{Sambe}. In fact, the basis $\left\{
|i^{\prime }(\tau )\rangle \right\} $ with $|i^{\prime }(\tau )\rangle
\equiv U^{+}(\tau )|i\rangle $ turns out to be the natural basis to express $%
|\Psi (\tau )\rangle $

\end{mathletters}
\begin{equation}
|\Psi (\tau )\rangle =C_{1}^{\prime }(\tau )|1^{\prime }(\tau )\rangle
+C_{2}^{\prime }(\tau )|2^{\prime }(\tau )\rangle +C_{3}^{\prime }(\tau
)|3^{\prime }(\tau )\rangle  \label{ec1.13}
\end{equation}
As this expression reflects, the dynamical evolution of the probability
amplitudes corresponding to the renormalized $\left\{ |i^{\prime }(\tau
)\rangle \right\} $ states is governed by the Hamiltonian $H^{\prime }$ of
Eq. (\ref{ec1.5}). Such Hamiltonian has the same form as the original
Hamiltonian (\ref{ec1.2}) in the limit $\Omega _{12}\rightarrow 0$ (in the
rotating wave approximation and in the frame rotating with the laser
frequency). Therefore, the theory is renormalizable in the sense that when
analyzed in terms of the $\left\{ |i^{\prime }(\tau )\rangle \right\} $
states, the nonperturbative effects of the radiation field on the dynamical
evolution of the system can be absorbed into the renormalized splitting $%
\Delta _{0}^{{\rm R}}$ and Rabi frequency $\Omega _{23}^{{\rm R}}$, in such
a way that the system evolves obeying the same Hamiltonian as that of the
weak-field regime in the rotating wave approximation. In fact, the general
solution (\ref{ec1.13}) is valid both in the (perturbative) weak-field
regime ($\Omega _{12}/\omega _{{\rm {L}}},\Delta _{0}/\omega _{{\rm {L}}}\ll
1$) and in the (nonperturbative) strong-field regime ($\Omega _{12}/\omega _{%
{\rm {L}}}\gtrsim 1$).

As Eq. (\ref{ec1.10}) shows, under the action of the coherent external
field, the $|1^{\prime }(\tau )\rangle $ state decouples and all the
population that is initially in state $|1\rangle $ becomes trapped in $%
|1^{\prime }(\tau )\rangle $. For a system prepared initially in its ground
state this implies, in particular, that the upper level $|3\rangle $ remains
always unpopulated. This occurs in spite of the fact that the initial
population oscillates very rapidly between the $|1\rangle $ and $|2\rangle $
levels, and the latter is directly coupled to $|3\rangle $ via a laser field
tuned to the $2\leftrightarrow 3$ transition. This is a coherent population
trapping phenomenon of nonperturbative origin.


\begin{figure}{\par\centering \resizebox{8.2cm}{!}{\rotatebox{0}
{\includegraphics{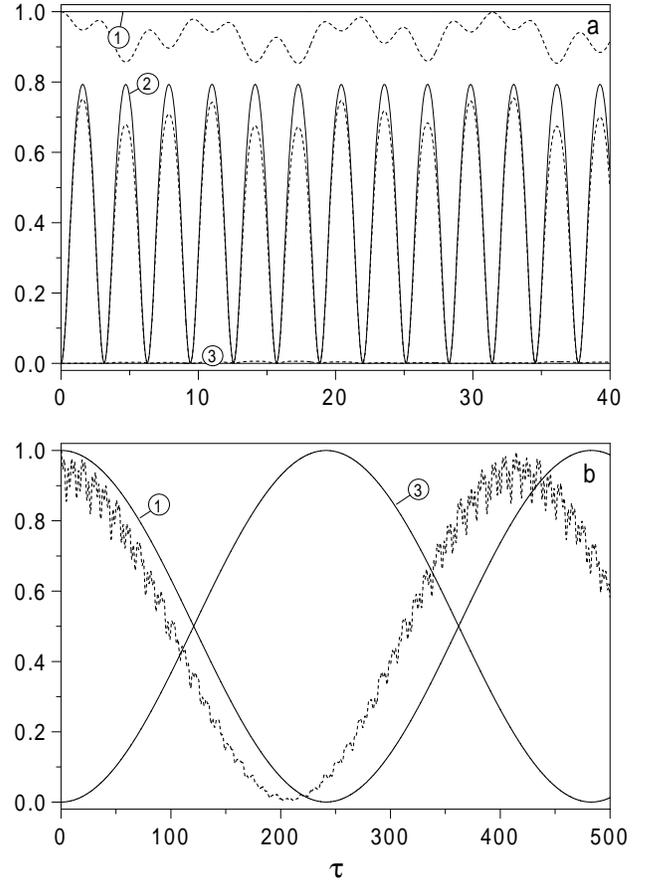}}} \par}
\caption{\small
Theoretical predictions for the same situations considered in Fig. 1. For
comparison purposes, along with the analytical results (solid lines) the
corresponding exact numerical results have been plotted again (dotted
lines). %
}\end{figure}

On the other hand, when the molecule is prepared in $\tau =0$ in the state $%
|2\rangle $, the population difference between the upper level and the lower
doublet oscillates in time as

\begin{equation}  \label{ec1.14}
W(\tau )=-\cos \left( \frac{\Omega ^{{\rm R}}}{\omega _{{\rm {L}}}}\tau
\right) -2\left( \frac{\delta ^{{\rm R}}}{\Omega ^{{\rm R}}}\right) ^2\sin
^2\left( \frac{\Omega ^{{\rm R}}}{2\omega _{{\rm {L}}}}\tau \right)
\end{equation}

Fig. 2 shows the theoretical predictions of our model for the same
situations considered in Fig. 1. Fig. 2a corresponds to an ammonia molecule
prepared initially in its ground state. In this case, the populations $\rho
_{ii}(\tau )$ of the molecular states $\left\{ |i\rangle \right\} $ are
predicted to be 
\begin{equation}
\rho _{22}(\tau )=\sin ^{2}\left( \frac{\Omega _{12}}{\omega _{{\rm {L}}}}%
\sin \tau \right) ,\;\;\;\;\rho _{33}(\tau )=0  \label{ec1.15}
\end{equation}
and $\rho _{11}(\tau )=1-\rho _{22}(\tau )$. Thus, the population of the
lower doublet, which under these circumstances coincides with that of the
renormalized state $|1^{\prime }(\tau )\rangle $, remains always equal to
one [curve (1)] and the upper level remains unpopulated [curve (3)], in good
agreement with the numerical results (dotted lines). Fig. 2a also compares
the analytical result for the population of level $|2\rangle $ [curve (2)]
with the corresponding exact numerical result.

On the other hand, Fig. 2b shows the evolution of the populations for a
molecule prepared at $\tau =0$ in state $|2\rangle $. In this case the
population of the lower doublet coincides with that of the renormalized
state $|2^{\prime }(\tau )\rangle $ and, according to our model, oscillates
in time as [curve (1)] 
\begin{equation}
\rho _{11}(\tau )+\rho _{22}(\tau )=\cos ^{2}\left( \frac{\Omega ^{{\rm R}}}{%
2\omega _{{\rm {L}}}}\tau \right) +\left( \frac{\delta ^{{\rm R}}}{\Omega ^{%
{\rm R}}}\right) ^{2}\sin ^{2}\left( \frac{\Omega ^{{\rm R}}}{2\omega _{{\rm 
{L}}}}\tau \right) ,  \label{ec1.16}
\end{equation}
while the population of the upper level behaves as $\rho _{33}(\tau )=1-\rho
_{11}(\tau )-\rho _{22}(\tau )$ [curve (3)]. These results are in good
qualitative agreement with the corresponding numerical results. The main
discrepancy between Figs. 2b and 1b comes from the rapid oscillatory
behavior of the upper level population. As already mentioned, it can be
shown that this discrepancy, which decreases as the laser intensity does,
originates from population transfers to levels adjacent to the $3^{+}$
level, which now have a more significant contribution. In fact, if the
numerical problem is restricted to the three levels most directly involved
then numerical and analytical results become indistinguishable on the scale
of the figures. Our three-level model already captures the essential
features of the system and enables us to understand the dominant behavior of
the populations in terms of a nonperturbative coherent population trapping
phenomenon. The upper level remains unpopulated when the molecule is
initially prepared in its ground state because such configuration
corresponds to an initial preparation in the trapping state $|1^{\prime
}(\tau )\rangle $.

Next, we will analyze the influence of dissipation on the coherent
population trapping phenomenon previously found. Spontaneous emission
effects can be conveniently incorporated by assuming that the upper level
decays radiatively into state $|2\rangle $ with an effective spontaneous
emission rate $\Gamma $. The dynamics of the system is now described in
terms of the density operator $\rho (t)$ which obeys the usual master
equation (in which we have retained nonsecular terms). By performing 
the unitary transformation (\ref{ec1.3}) one obtains
a transformed master equation for the density operator $\rho ^{\prime
}(t)=U(t)\rho (t)U^{+}(t)$, which, within the range of validity of our
model, leads to the following equations of motion governing the time
evolution of populations and coherences:

\begin{mathletters}
\begin{equation}  \label{ec1.17}
\dot \rho _{11}^{\prime }=\frac \Gamma 2\left( 1-\Lambda _0\right) \rho
_{33}^{\prime }
\end{equation}

\begin{equation}
\dot{\rho}_{22}^{\prime }=i\Omega _{23}^{{\rm R}}\left( \rho _{32}^{\prime
}-\rho _{23}^{\prime }\right) +\frac{\Gamma }{2}\left( 1+\Lambda _{0}\right)
\rho _{33}^{\prime }  \label{ec1.18}
\end{equation}

\begin{equation}
\dot{\rho}_{12}^{\prime }=i\Delta _{0}^{{\rm R}}\rho _{12}^{\prime }-i\Omega
_{23}^{{\rm R}}\rho _{13}^{\prime }  \label{ec1.19}
\end{equation}

\begin{equation}
\dot{\rho}_{13}^{\prime }=i\left( \delta ^{{\rm R}}+\Delta _{0}^{{\rm R}%
}\right) \rho _{13}^{\prime }-i\Omega _{23}^{{\rm R}}\rho _{12}^{\prime }-%
\frac{\Gamma }{2}\left( \rho _{13}^{\prime }-\frac{1}{2}\Lambda _{2}\rho
_{31}^{\prime }\right)  \label{ec1.20}
\end{equation}

\begin{equation}
\dot{\rho}_{23}^{\prime }=i\delta ^{{\rm R}}\rho _{23}^{\prime }-i\Omega
_{23}^{{\rm R}}\left( \rho _{22}^{\prime }-\rho _{33}^{\prime }\right) -%
\frac{\Gamma }{2}\left( \rho _{23}^{\prime }-\frac{1}{2}\Lambda _{2}\rho
_{32}^{\prime }\right)  \label{ec1.21}
\end{equation}
with $\Lambda _{n}\equiv J_{n}\left( 2\Omega _{12}/\omega _{{\rm {L}}%
}\right) $ $(n=0,2);$ $\rho _{ij}^{\prime }=\langle i\left| \rho ^{\prime
}(t)\right| j\rangle =\langle i^{\prime }(t)\left| \rho (t)\right| j^{\prime
}(t)\rangle $; and $\rho _{ji}^{\prime }=\rho _{ij}^{\prime *}$.

As Eq. (\ref{ec1.17}) reflects, whenever the Rabi frequency $\Omega _{12}$
coupling the two lower-lying states is nonzero, the upper state $|3\rangle $
remains unpopulated in the steady state regardless of the initial
preparation. As a consequence, in the steady state the molecular population
becomes trapped in the lower doublet and the fluorescence from level $%
|3\rangle $ vanishes. This behavior, which is in sharp contrast with the
well-known behavior of the system in the $\Omega _{12}=0$ limit, is typical
of systems exhibiting coherent population trapping and has its origin in
quantum interferences involving the two lower-lying levels \cite{Ari1}.

It is not hard to see from the above equations that, for arbitrary external
fields (such that $\Omega _{23}^{{\rm R}}\neq 0$), the steady-state
population of $|2^{\prime }(t)\rangle $ also vanishes so that all of the
population becomes trapped in the steady state in $|1^{\prime }(t)\rangle $
irrespective of the initial preparation.

In conclusion, we have shown that coherent population trapping can occur in
the nonperturbative regime and have proposed an analytically solvable
nonperturbative three-level model which enables us to understand the
observed phenomenon. 
Although we have presented results for only one field intensity,
essentially the same behavior, in good agreement with our analytic
model, occurs in the parameter range $ 0.1 \pi \lesssim \Omega _{12}/
\omega _{{\rm {L}}} \lesssim 0.5 \pi$.
A detailed account of the model will be given elsewhere.

This work has been supported by MCYT and FEDER under Grant No. BFM2001-3343.

\end{mathletters}


\begin{references}
\bibitem{Hund}  F. Hund, Z. Phys. {\bf 43}, 803 (1927).

\bibitem{Holt1}  M. Holthaus and D. Hone, Phys. Rev. B {\bf 47}, 6499 (1993).

\bibitem{Oppen}  A. Oppenl\"{a}nder, Ch. Rambaud, H. P. Trommsdorff, and J.
C. Vial, Phys. Rev. Lett. {\bf 63}, 1432 (1989).

\bibitem{Fic1}  Z. Ficek and H. S. Freedhoff, in {\em Progress in Optics,}
edited by E. Wolf (Elsevier, Amsterdam, 2000), p. 389.

\bibitem{Ari1}  G. Alzetta, A. Gozzini, L. Moi and G. Orriols, Nuovo
Cimento B {\bf 36}, 5 (1976); E. Arimondo and G. Orriols, Lett. Nuovo
Cimento {\bf 17}, 333 (1976);
E. Arimondo, in {\em Progress in Optics,} edited by E. Wolf
(Elsevier, Amsterdam, 1996), p. 257 and references therein.

\bibitem{Harr1}  S. E. Harris, Phys. Today {\bf 50}, No. 7, 36 (1997).

\bibitem{Koch1}  O. Kocharovskaya and Ya. I. Khanin, JETP Lett. {\bf 48}, 
630 (1988); S. E. Harris, Phys. Rev. Lett. {\bf 62}, 1033 (1989); 
A. Imamoglu, Phys. Rev. A {\bf 40}, 2835 (1989); M. O. Scully 
{\em et al.,} Phys. Rev. Lett. {\bf 62}, 2813 (1989); G. S. Agarwal, 
Phys. Rev. A {\bf 44}, R28 (1991).

\bibitem{Gros1}  F. Grossmann, T. Dittrich, P. Jung, and P. H\"{a}nggi,
Phys. Rev. Lett. {\bf 67}, 516 (1991); M. Grifoni and P. H\"{a}nggi, Phys.
Rep. {\bf 304}, 229 (1998).

\bibitem{Sambe}  H. Sambe, Phys. Rev. A {\bf 7}, 2203 (1973).
\end{references}
\end{document}